# DETERMINATION OF SENSITIVITY OF SEMICONDUCTOR DETECTORS OF GAMMA-RADIATION

## A.A. Zakharchenko

*National Science Center "Kharkov Institute of Physics and Technology"*
*1, Akademicheskaya St., 61108 Kharkov, Ukraine*
*e-mail: az@kipt.kharkov.ua*

Properties of response functions of room temperature gamma-radiation detectors based on wide band-gap semiconductors are researched using Monte-Carlo method. It is shown that approximate formulas which connect detector sensitivity with absorbed energy of monochromatic radiation in the energy range between 0.06 and 3 MeV can be obtained for some kinds of semiconductors. We determined gamma-quantum energy regions and detector thicknesses where obtained approximate formulas are correct. Regions of maximum error of approximate formulas are also determined.

The measurement of energy dependence of gamma-radiation detector sensitivity is one of the laborious problems at investigation of room-temperature semiconductor detectors (SCD) which work in the pulse counting mode or pulse amplitude analysis mode. More than 10 certified gamma-radiation sources are required for measurement of SCD sensitivity in the energy range between 0.03 and 3 MeV. Most of these sources are produced from short-lived radioactive isotopes that have half-live about 1…2 years [1]. The lack of uniformity of electrophysical characteristics of room-temperature SCD [2] is a reason of considerable spread of absolute value of sensitivity δ and reason of necessity of detailed measurements of energy dependence $\delta(E_\gamma)$ for each detector. Through instability of SCD contact characteristics [3] it is necessary to repeat these measurements of $\delta(E_\gamma)$ periodically in order to have an opportunity to agree data obtained at different time periods.

Moreover isotopic gamma-radiation sources are not monochromatic. Efficiency of low energy (less than 60 keV) gamma-quantum registration with SCD is no less than 80%. At the same time the efficiency of relatively high energy (more than 500 keV) gamma-quantum registration with SCD does not exceed 5–7%. When these both energy lines are present in the spectrum of the same source the Compton scattering region of high energy gamma-quanta is distorted by pulses from fully absorbed low energy quanta.

The Monte-Carlo simulation of SCD response can be used to overcome these difficulties but to obtain coincident results it is necessary to know control parameters of detector model with adjusted precision that is not always possible [2].

In this work we present results of numerical experiments to study the gamma-radiation semiconductor detector response functions $\delta(E_\gamma)$ for $HgI_2$, TlBr and CdZnTe semiconductor compound. Approximation formulas for calculation of energy dependence $\delta(E_\gamma)$ of SCD sensitivity are obtained based on the analysis properties of detector response functions. These formulas can be efficient for detector sensitivity calculation in the whole range of gamma-quantum energy between 0.06 MeV and 3 MeV using experimental measurements at the few energies. Maximum disagreement between approximation formula data and detailed Monte-Carlo calculation are evaluated.

**FACTORS AFFECTING MEASUREMENTS OF ENERGY DEPENDENCE OF SCD SENSITIVITY**

The sensitivity δ of semiconductor detector which used to detect single gamma-quanta (discrete sensitivity) is defined as ratio of pulse number $N$ to unit of radiation dose [4]. Energy dependence of cross-sections of gamma-quantum interaction with detector material is a basic factor which defines the behavior of sensitivity $\delta(E_\gamma)$ with change of gamma-quantum energy. The total attenuation coefficient of gamma-quantum flow (μ) in the semiconductors is changed on four orders of magnitude in the energy range between 10 keV and 1 MeV (Fig. 1). The final form of $\delta(E_\gamma)$ (Fig. 2) depends on charge collection efficiency (*CCE*) of nonequilibrium carriers that are created at full or partial absorption of gamma-quantum energy inside detector material.

Fig. 2 shows examples of experimentally measured dependencies of $\delta(E_\gamma)$. For convenience both curves are normalized to detector sensitivity $\delta_{0,1}$ which is measured at the energy $E_\gamma$ = 0.1 MeV. CdTe detector with thickness of 2.8 mm is investigated in the Ref. [6] so $\delta_{CdTe}(E_\gamma)$ is measured in the energy range between 0.06 and 1.2 MeV. The energy dependence of CdTe detector sensitivity $\delta_{CdTe}(E_\gamma)$ is built according to measurement results with 9 gamma-radiation sources. Thickness of $HgI_2$ detector was about 0.5 mm and the measured energy dependence $\delta_{HgI_2}(E_\gamma)$ is restricted within range between 0.005 and 0.3 MeV [7]. Thickness of $HgI_2$ detector is a reason of low efficiency of gamma-quantum registration in the energy range more than 0.2 MeV and observable sharp loss of detector sensitivity (Fig. 2).

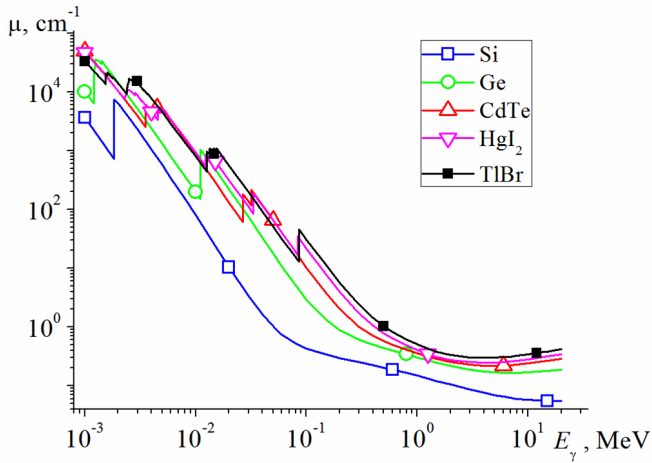
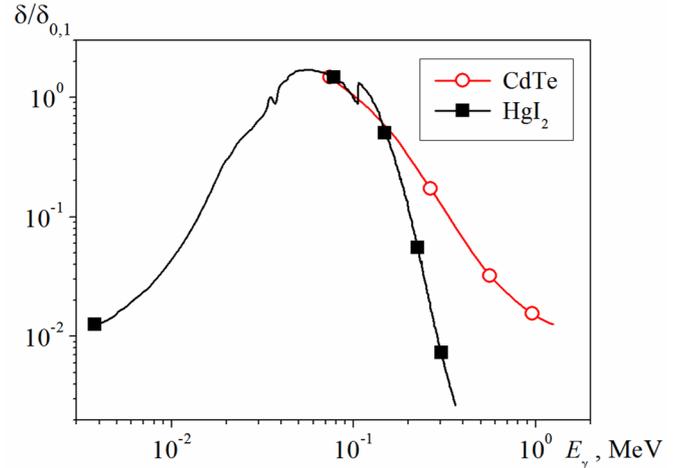

Fig. 1. The linear attenuation coefficient of gamma-quantum flow in the some semiconductors [5].

Fig. 2. The dependence of sensitivity vs gamma-quantum energy for CdTe [6] and HgI$_2$ [7] detectors.

*CCE* is defined by great number of factors: coordinates of gamma-quantum interaction inside detector, mobility and lifetime of charge carriers (electrons and holes) and internal electric field intensity inside the detector [8, 9]. When detector with planar contacts and uniform distribution of internal electric field registers interaction between single gamma-quantum and detector material that *CCE* function has the simple form [8]:

$$CCE(x,d,\mu_e,\tau_e,\mu_h,\tau_h,U) = \frac{Q_{ind}}{Q_{tot}} = \frac{\mu_e \tau_e U}{d^2}\left(1 - e^{\frac{-(d-x)d}{\mu_e \tau_e U}}\right) + \frac{\mu_h \tau_h U}{d^2}\left(1 - e^{\frac{-xd}{\mu_h \tau_h U}}\right), \quad (1)$$

where $Q_{ind}$ – charge that is induced on the detector contacts; $Q_{tot}$ – average charge that is produced at absorption of the energy $E$ in the detector; $d$ – detector thickness; $x$ – depth of gamma-quantum interaction with detector matter ($0 < x < d$); $\mu_{e,h}$ ($\tau_{e,h}$) – average mobility (mean lifetime) of electrons and holes; $U$ – SCD bias.

Considerable variations of mobility ($\mu_{e,h}$) and mean lifetime ($\tau_{e,h}$) of nonequilibrium charge carriers are characteristic feature of all semiconductor materials which can be used for producing room-temperature gamma-radiation detectors [2]. The reasons of nonuniformity of electrophysical detector characteristics were not finally determined. For example, tellurium precipitate formation in the single crystal growing process can be defined as one of the factors that have influence on the characteristics CdZnTe detectors [10]. The sizes of these tellurium precipitates are changed in the range from about 1–2 μm to several hundred microns [11]. A great number of vacancies are formed near the bounds of Te precipitate. They form complexes of defects with Te atoms and trap holes at energetic levels located near top of the valence band [12]. These traps can be reason of the dark current increasing. As the traps are nonuniformly distributed in the crystal [10, 11] that variations of dark currents and, correspondingly, variations of intrinsic noises are unexpected from one detector to another.

The Refs. [10, 11] also showed that considerable variety of defects (point, lengthy, volumetric) is typical for as-grown CdZnTe ingots. In the aggregate these defects can essentially decrease charge carrier mobility and life time in the detector and at the same time they also can increase dark currents. The negative consequence of these effects is reduction of limiting work detector bias. The results of Ref. [13] show that if at selected SCD bias the intensity of internal electric field is less than 150 V/mm that planar CdZnTe detectors have inactive areas with null charge collection efficiency *CCE*. If the internal electric field intensity is less than 100 V/mm that *CCE* in the CdZnTe detectors can be decreased down to 4–5 times relatively maximum for specific detector.

One more factor influenced on the gamma-radiation SCD sensitivity is an polarization effect. In the general case the term «polarization» is used for identification of uncontrolled time changes of nuclear radiation detector characteristics after bias set up [14]. These changes can appear at the time distances from about a few minutes to a few weeks. Good known demonstrations of such changes are time variations of the energy resolution and gamma-quantum registration efficiency. The change of the registration efficiency directly influence on SCD sensitivity.

In the next part Monte-Carlo simulation of SCD response is used for illustration of influence of some above-listed processes on detector sensitivity.

**SIMULATION OF RESPONSE FUNCTION OF ROOM-TEMPERATURE SCD**

The model calculation of response functions of room-temperature SCD was originally developed and tested for CdTe and CdZnTe detectors equipped planar contacts [15, 16]. Subsequently, this model has been applied for analysis of characteristics of other wide band-gap semiconductor materials [17, 18]. The universal EGSnrc code for the Monte-Carlo simulation of the passage of photons, electrons and positrons through matter [19] was used for computation. In spite of interaction of gamma-quanta and charged particles with detector matter the model [15] allow to take into account the influence of noises and losses of nonequilibrium charge on the output signal amplitude.

Fig. 3 and 4 show calculated response function of $HgI_2$ detector equipped planar contacts on gamma-quanta with energy $E_\gamma = 662$ keV ($^{137}$Cs source) and its transform in the measuring channel. Ordinate is defined as the ratio of pulse counts $N_i$ in $i$ channel of the simulated analog-digital converter (ADC) to total pulse counts $N_{tot} = \sum_i N_i$. Detector parameters correspond to data Ref. [20]: sizes of $1\times1\times1$ cm$^3$, charge transport characteristics – $(\mu\tau)_e = 5\times10^{-3}$ cm$^2$/V, $(\mu\tau)_h = 3\times10^{-5}$ cm$^2$/V, SCD bias $U = 2.5$ kV. Fig. 3 demonstrates processes of gamma-quantum energy absorption and nonequilibrium charge production in the detector. The investigated detector has sufficiently high probability of the complete absorption (photoeffect) of gamma-quanta with energy $E_\gamma = 662$ keV. Simulation shows that about fifty percent of the interacted quanta are fully absorbed. The efficiency of gamma-quanta registration for energy $E_\gamma = 662$ keV is about 42% and so absolute probability of scattering in the photopeak is about 21%. It means that less than 25% of input gamma-quantum flow is fully absorbed in $HgI_2$ at the length 1 cm. Escape peaks corresponded to characteristic gamma-quanta of the mercury and iodine which have left the detector volume are clearly visible on the Compton valley ($0.45 < E < 0.65$ MeV). The ratio of photopeak amplitude $E_\gamma$ to the average pulse amplitude in the Compton continuum region ($0 < E < 0.45$ MeV) exceeds two orders of magnitude. Small peak spreading is connected with fast electron energy losses due to the generation of lattice vibrations (losses up to 5%).

Fig. 4 shows changes of response function of $HgI_2$ detector (Fig. 3) after the output of the measuring channel. A main part of pulse amplitudes is shifted to the Compton continuum region. The escape peaks are almost completely spreaded. The ratio of the photopeak amplitude $E_\gamma$ to the average amplitude in the Compton continuum region (i.e. where $0 < E < 0.45$ MeV) is less than 1.3. In whole the response function (Fig. 4) has satisfactory agreement with calculation in Ref. [20].

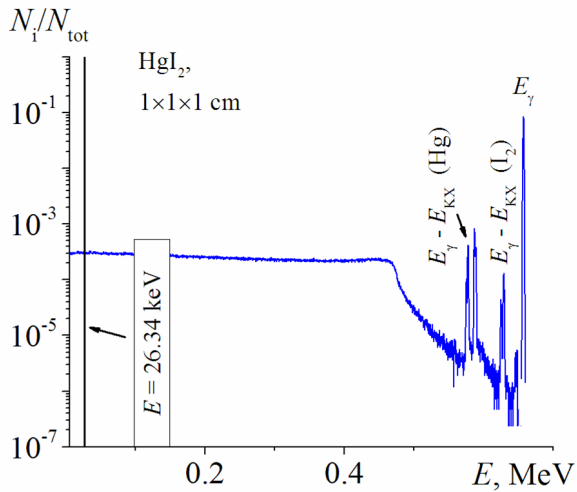

Fig. 3. The response function of $HgI_2$ detector on gamma-quanta with energy 662 keV without taking into consideration the charge collection and noises.

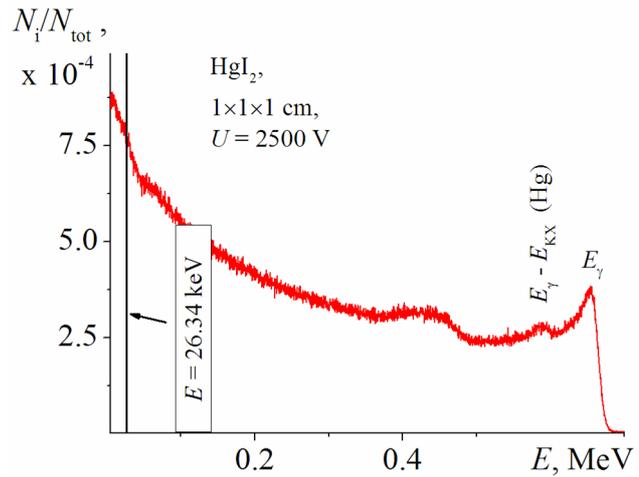

Fig. 4. The response of $HgI_2$ detector after the output of the measuring channel corresponded to Ref. [20].

The total number of pulses $N_{tot}$ is the same in both cases (Fig. 3 and Fig. 4). However it is necessary to notice that in the real measurements the initial part of region of small amplitudes is not taken into account by reason of its overlapping with noise region. If the discrimination threshold is specified as 26.34 keV (it is one of the gamma-lines of $^{241}$Am source) that sensitivity value $\delta_3$ is about 1249.4 (Fig. 3) and $\delta_4$ is about 1163.9 pulse/μR (Fig. 4), respectively. In other words experimentally measured value of sensitivity does not exceed 93% from theoretically possible value (ratio $\delta_4/\delta_3$ is about 0.93). It is an acceptable result.

Unfortunately, mentioned value of difference is not final. Results of simulation (Fig. 4) are satisfactorily conformed to calculations that presented in the Ref. [20] but the shape of the showed experimental $^{137}$Cs spectrum differs from the shape of calculated spectrum. The most important distinctions are the photopeak lack and higher value of experimental ratio $N_i/N_{tot}$ at the initial energy range in comparison with simulation. The variation of the model parameters showed that we can achieve such spectral distribution if the intensity of internal electric field inside the detector is much less than expected value that is $U/d = 2500$ V/cm for planar detectors. It can be a consequence of the distortion of the internal electric fields near numerous growth defects of semiconductors as in Ref. [13, 21]. This case can be reproduced with the used model at lower bias $U$ (Fig. 5). From Fig. 5 it follows that the photopeak is absent as it occurs for the real spectrum of $^{137}$Cs source that was presented in Ref. [20]. The pulse number in the initial channels exceeds the average value of pulse number in the Compton continuum up to one order of magnitude.

Fig. 6 shows change *CCE(x)* corresponding to reduction of the internal electric field intensity inside the detector. At the same time the detector thickness-averaged collection efficiency *CCE* is decreased from 57.6% to 42.9%. It leads to complete degradation of the 662 keV photopeak (Fig. 5).

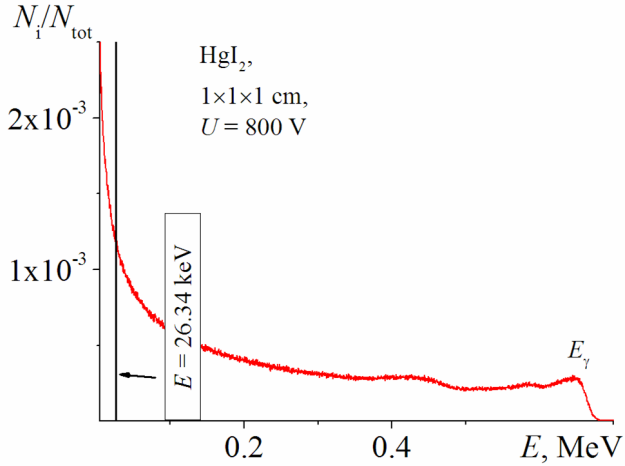
Fig. 5. The change of HgI$_2$ detector response at the bias reduction.

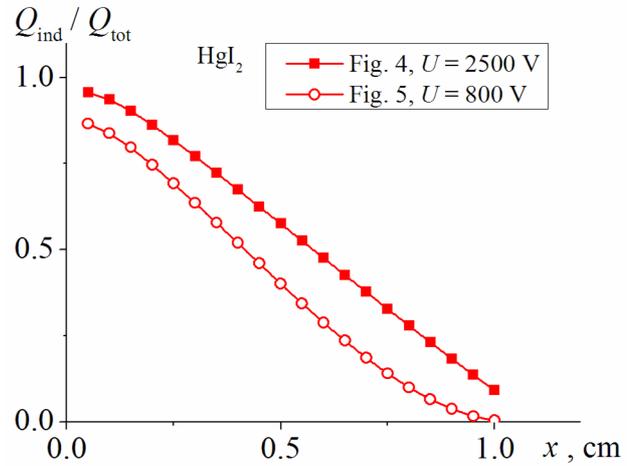
Fig. 6. The change of the charge collection efficiency CCE(x) in the HgI$_2$ detector.

Calculated sensitivity corresponding to condition of simulation that showed on Fig. 5 is $\delta_5 \approx 1033.4$ pulse/μR and ratio ($\delta_5/\delta_3$) is about 0.83. With the noise level increasing the discrimination threshold needs to be specified up to 60 keV and more. As the result the difference between measured sensitivity and maximum possible value of sensitivity can be even worse than mentioned ratio ($\delta_5/\delta_3$).

This example presents mediated influence of the defects on SCD sensitivity and problems of recovery of the energy dependence of SCD sensitivity $\delta(E_\gamma)$ using the Monte-Carlo method. The change of value of the internal electric field intensity is one of the ways to make agree between simulation and experiment data. The similar result for this HgI$_2$ detector can be obtained by changing the ratio between $(\mu\tau)_e$ and $(\mu\tau)_h$. Additional experimental measurements are necessary to choose the variant of matching.

Approximate formulas with several (2 or 3) fit parameters which allow to calculate the sensitivity value $\delta(E_\gamma)$ for the wide gamma-quantum energy range (for example, 0.04…3 MeV) can be useful at the limited set of the experimental data.

Next part presents the results of the simulation of the gamma-radiation SCD response functions and the calculation of the function $\delta(E_\gamma)$. The group of the semiconductor materials which are intensively researched for the past years [22] was chosen for this investigation. For comparison simulation of the cooled germanium detector was run, too. Table 1 contains the parameters of the simulated detectors.

Table 1.
The characteristics of the simulated gamma-radiation detectors

| No | Detector | Sizes | $(\mu\tau)_e$, cm$^2$/V | $(\mu\tau)_h$, cm$^2$/V | U, V | Refs |
|---|---|---|---|---|---|---|
| 1 | HgI$_2$ | 25×25×2.89 mm$^3$ | | | 2800 | [23] |
| 2 | TlBr | 7 mm$^2$ × 1 mm | 5×10$^{-4}$ | 2×10$^{-6}$ | 200 | [24] |
| 3 | CdZnTe | 6×6×3 mm$^3$ | 1.1×10$^{-4}$ | 1.3×10$^{-5}$ | 300 | [15] |
| 4 | CdZnTe | 4×4×2.5 mm$^3$ | 4×10$^{-3}$ | 2×10$^{-5}$ | 400 | [25] |
| 5 | Ge | 1000 mm$^2$×13 mm | 1 | 1 | 1000 | [26] |

The intensity of the simulated gamma-quantum flux is the same (10$^3$ quantum/s) for whole considered energy range between 0.04 and 3 MeV. It is supposed that values of noise characteristics of the simulated measuring channel are like that: equivalent noise charge is 300 e$^-$ (charge units) and leakage current (or dark detector current) is 5 nA. Thus total noise level (detector + measuring channel) is about 400 e$^-$ or less than 2 keV. The discrimination threshold for calculation of the sensitivity has been specified as 26.34 keV because it is proper to set up in the experimental measurements using $^{241}$Am source. The shaping time is 1 μs. The dead time of the simulated analog-digital converter is 100 μs.

**RESULTS AND DISCUSSION**

Fig. 7 shows the results of simulation of the sensitivity dependence $\delta(E_\gamma)$ from the energy of the registered gamma-quanta for CdZnTe detectors no. 3 and 4 (Table 1). Previously it was demonstrated that the simulation data satisfactorily agree with results of the metrological measurements [15]. The comparison between Fig. 2 and Fig. 7 confirms the trend of the $\delta(E_\gamma)$ curves. The smaller volume of detector no. 4 as compared with detector no. 3 is the cause of lower values of its sensitivity $\delta(E_\gamma)$.

The analysis of the functional dependence $\delta(E_\gamma)$ for CdZnTe detectors is simplified if the inverse function $1/\delta = f_0(E_\gamma)$ is used. Fig. 8 shows that the functions $1/\delta$ are nearly linear in the energy range $E_\gamma$ above 0.2 MeV. It allows to suppose that the dependence $\delta(E_\gamma)$ at $E_\gamma \geq 0.2$ MeV can be restored using the function $1/\delta_{fit} = a_0 \times E_\gamma + b_0$. The Table 2 contains the fitting data obtained with the method of least squares at the different initial energies $E_0$ included in the fitting region.

Data from Table 2 show that the coefficients of the fitting curve are weakly changed even if the nonlinear region between 0.06 and 0.2 MeV is also fitted. Fig. 9 shows the relative difference $\eta$ between the inverse sensitivity values of CdZnTe detector no. 4 obtained using the simulation data $1/\delta_{simul}$ and using the next expression for linear function $1/\delta_{fit} = a_0 \times E_\gamma + b_0$:

$$\eta = \frac{\frac{1}{\delta_{fit}} - \frac{1}{\delta_{simul}}}{\frac{1}{\delta_{simul}}} \times 100\%. \qquad (2)$$

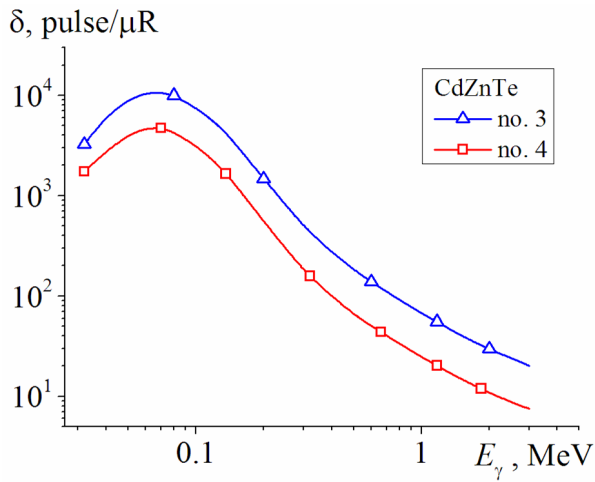
Fig. 7. The dependence of sensitivity vs the energy of registered gamma-quanta for CdZnTe detectors.

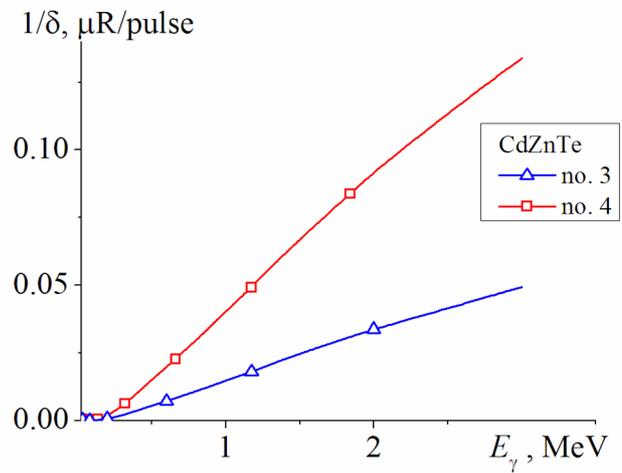
Fig. 8. The dependence $1/\delta = f_0(E_\gamma)$ for CdZnTe detectors.

Table 2.
The parameters of the linear fitting of $1/\delta(E_\gamma)$ for CdZnTe detectors (Fig. 8)

| Initial point, MeV | no. 3 | | no. 4 | |
|---|---|---|---|---|
| | $a_0$ | $b_0$ | $a_0$ | $b_0$ |
| 0.06 | 0.0175±0.0002 | −0.0024±0.0002 | 0.0476±0.0005 | −0.0064±0.0006 |
| 0.1 | 0.0177±0.0002 | −0.0027±0.0002 | 0.0482±0.0005 | −0.0074±0.0006 |
| 0.2 | 0.0180±0.0002 | −0.0032±0.0003 | 0.0489±0.0005 | −0.0086±0.0007 |

The difference curves for CdZnTe detector no. 3 do not demonstrate essentially distinctions in comparison with detector no. 4. Fig. 9 shows that the satisfactory fit between the simulation and the fitting data in the energy range $E_\gamma$ above 0.5 MeV is obtained even if the initial point of fitting specified as $E_0 = 60$ keV. It means that high-energy part of the $\delta(E_\gamma)$ dependence for CdZnTe detectors can be restored with satisfactory accuracy using measurement results with three sources: $^{241}$Am ($E_\gamma$ equals 60 keV), $^{137}$Cs ($E_\gamma$ equals 662 keV), $^{60}$Co (doublet 1173 and 1333 keV). Ukrainian national primary standard of the units of exposure dose and exposure dose rate of X- and γ-radiation contains these three sources. Fig. 10 shows the examples of the spectra of $^{137}$Cs and $^{60}$Co sources calculated for CdZnTe detector no. 3.

The simulation of the spectra of $^{137}$Cs and $^{60}$Co sources presents above mentioned problems of the experimental sensitivity measurement connected with the non-monochromatic of gamma-radiation. Although the relative intensity of low-energy multiplet of $^{137}$Cs source centered around 32.2 keV is about 6% (at the same time the intensity of 662-keV line produced by the $^{137}$Cs source is 85%) but its contribution in the value of $\delta$ can be considerable. Thus, if the discrimination threshold is specified as 26.34 keV that the value of sensitivity of CdZnTe detector no. 3 calculated using the simulation of the $^{137}$Cs source (Fig. 10) is 166.1 pulse/μR whereas for the 662-keV monochromatic line sensitivity is 120.4 pulse/μR.

The problem of the non-monochromaticity for the $^{137}$Cs source can be solved using the thin metallic filter absorbed low energy gamma-quanta or increasing the discrimination threshold up to 40–45 keV. However the separation of the neighboring lines with the same intensity is not possible for the $^{60}$Co source. The calculated value of sensitivity for the $^{60}$Co source is 48.5 pulse/μR (Fig. 10). It corresponds to gamma-quantum energy about 1.28 MeV. The simulation predicts the sensitivity values 55.4 (1173 keV) and 46.8 (1333 keV) pulse/μR for monochromatic lines that correspond to $^{60}$Co source. In other words at the experimental measurements of $\delta(E_\gamma)$ the characteristics of the real $^{60}$Co source need to be recalculated for the equivalent source of gamma-quanta with energy 1.28 MeV.

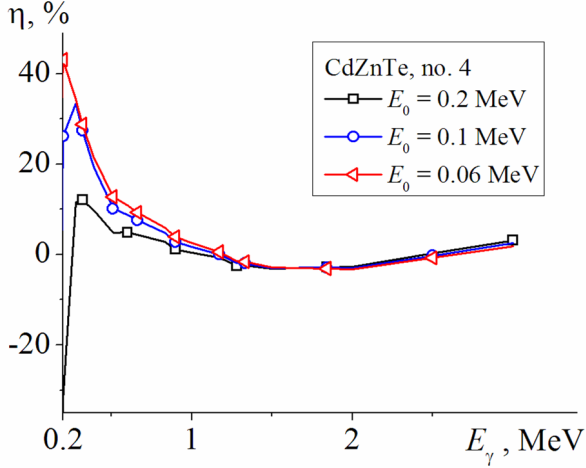

Fig. 9. The difference between the inverse sensitivity values for the simulation and fitting (Fig. 8).

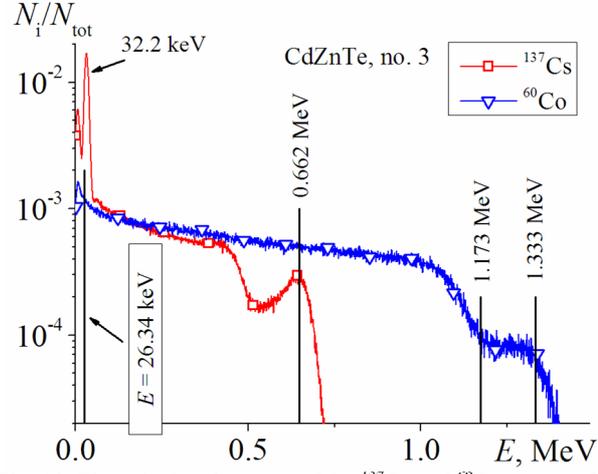

Fig. 10. The calculated spectra of the $^{137}$Cs and $^{60}$Co sources for CdZnTe detector.

As Fig. 11 shows the satisfactory approximation $1/\delta = f_0(E_\gamma)$ in the energy range $E_\gamma$ above 0.2 MeV can be obtained if function $f_0(E_\gamma)$ has either quintic polynomial form or form of

$$\frac{1}{\delta} = \frac{abE_\gamma^{1-c}}{1+bE_\gamma^{1-c}}, \qquad (3)$$

where $a$, $b$ and $c$ – – the fitting parameters of the extended Langmuir model – ELM.

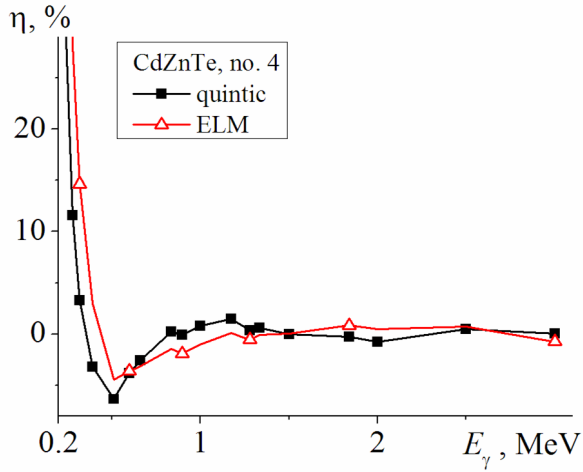

Fig. 11. The difference between inverse sensitivity values for the simulation and different kinds of the fitting.

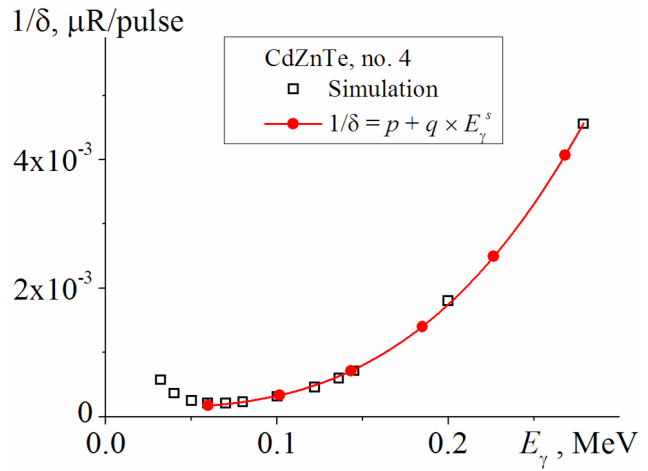

Fig. 12. The fitting of the dependence $1/\delta$ using the power function for CdZnTe detector ($p$, $q$, $s$ – fitting parameters).

The number of the experimental points which are necessary for the reliable determination of the polynomial coefficients can exceed the number of the usually accomplished measurements of detector sensitivity. Moreover the difference between the both approximate formulas (the quintic or ELM) and the data of the sensitivity simulation of the CdZnTe detectors in the energy range $E_\gamma$ below 0.2 MeV is too large. Fig. 12 shows that power function can satisfactorily fit $1/\delta$ in the energy range between 0.06 and 0.3 MeV. The uniform distribution of the section of fitting between 0.06 and 0.1 MeV by experimental data is desirable for the reliable determination of the power function coefficients. However in practice either one point (60 keV, $^{241}$Am) is presented in the energy range between 0.06 and 0.1 MeV [6] or the measurements are grouped around one or two energies [1, 7]. As Fig. 12 and Fig. 13 show the satisfactory fit between approximate and simulated values of $1/\delta(E_\gamma)$ in the whole energy range between 0.06 and 3 MeV can be obtained with increasing the number of the points in the energy range $E_\gamma$ from 0.06 to 0.1 MeV. The maximum error of the approximate formulas for the calculation of $1/\delta(E_\gamma)$ for CdZnTe detectors is found in the energy range $E_\gamma$ between 0.3 and 0.4 MeV and can exceed up to +10% (Fig. 13). Consequently the maximum error of the calculation of $\delta(E_\gamma)$ using the approximate formulas is −10% in relation to the Monte-Carlo $\delta(E_\gamma)$ simulation data. Thus the obtained detector sensitivity value will be found underestimated.

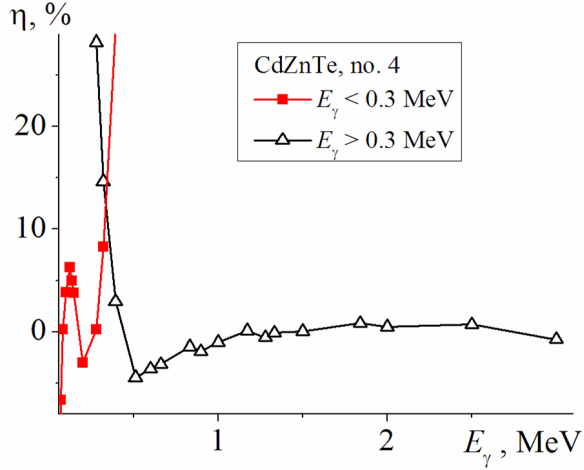

Fig. 13. The difference between the approximate formulas (4) and the Monte-Carlo simulation data for CdZnTe detectors.

Consequently for CdZnTe SCD the approximate formulas for the calculation of $\delta(E_\gamma)$ are expressed as

$$\frac{1}{\delta} = \begin{cases} p + qE_\gamma^s, & E_\gamma < 0.3 \text{ MeV, (a)} \\ \dfrac{abE_\gamma^{1-c}}{1+bE_\gamma^{1-c}}, & E_\gamma > 0.3 \text{ MeV, (b)} \end{cases}. \quad (4)$$

The maximum difference between $\delta(E_\gamma)$ values calculated using the formulas (4) and the detailed Monte-Carlo simulation data is 10%.

The subsequent simulation of the sensitivity $\delta(E_\gamma)$ for other semiconductor materials ($HgI_2$, TlBr) was run for verification of the universality of formulas (4) obtained for CdZnTe detectors. Fig. 14 shows the dependence of $1/\delta$ for $HgI_2$ detector (no. 1, Table 1) and its fitting using the formula (4b). Fig. 15 shows the difference between the calculation by formulas (4) and the data of the simulation of sensitivity $HgI_2$ detector in the whole energy range between 0.06 and 3 MeV.

Simulation demonstrates that the linear dependence of $1/\delta$ is disturbed for $HgI_2$ detectors at energy $E_\gamma$ above 2 MeV (Fig. 14). It means that the method of the reconstruction of the high-energy part of $\delta(E_\gamma)$ using three measurements can not be used for $HgI_2$ detectors although it has been worked for CdZnTe detectors (Fig. 8). At the same time the difference between the results of the $1/\delta$ calculation used the approximate formulas (4) and the Monte-Carlo simulation data for $HgI_2$ detectors (Fig. 15) is similar with that for CdZnTe detectors (Fig. 13). The maximum error of the approximate formulas (4) for $HgI_2$ detectors also was found in the energy range $E_\gamma$ between 0.3 and 0.4 MeV and can exceed up to +15%. It is slightly worse than for CdZnTe detectors.

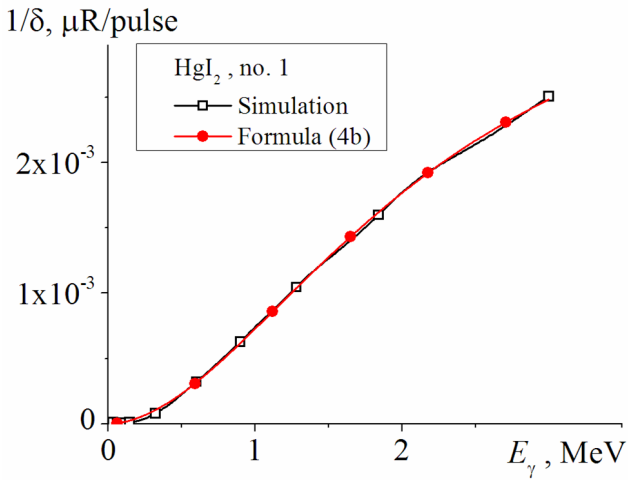

Fig. 14. The $1/\delta$ dependence for $HgI_2$ detector.

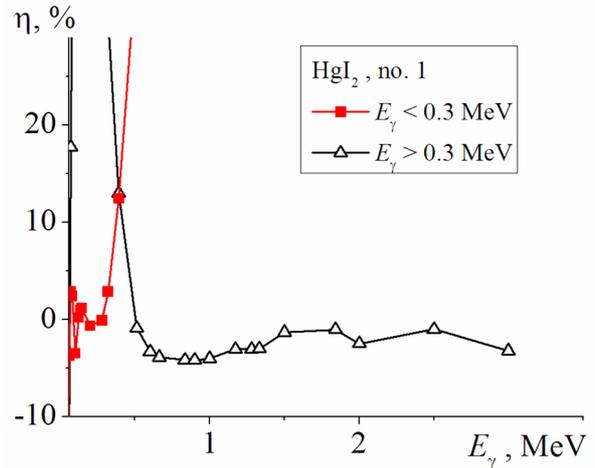

Fig. 15. The difference between the approximate formulas (4) for calculation of $1/\delta$ and the Monte-Carlo simulation data for $HgI_2$ detector.

Fig. 16 shows the $1/\delta$ dependence for TlBr detectors (no. 2, Table 2) in the energy range $E_\gamma$ below 0.3 MeV and its fitting by the formula (4a). Fig. 17 presents the full $1/\delta$ dependence and its fitting by the formula (4b). As we can see the $1/\delta$ dependence linearity is absent for TlBr detector at $E_\gamma$ above 1.5 MeV. Thus the method of the reconstruction of the high-energy part of $\delta(E_\gamma)$ using three measurements can not be used for TlBr detectors, too (similarly $HgI_2$).

Fig. 18 presents the difference between the $1/\delta$ calculation used the formulas (4) and the data of the TlBr detector simulation in the energy range from 0.06 to 3 MeV. The obtained results show that the more essential difference between the approximate formulas (4) and the detailed Monte-Carlo calculation is typical for TlBr in the gamma-quantum energy range from 0.3 to 0.4 MeV. The maximum error of the $1/\delta$ calculation exceeds up to +25%. Consequently the maximum error of the calculation $\delta(E_\gamma = 0.3\ldots0.4$ MeV) is −25%. Moreover the calculation error $\delta(E_\gamma = 0.07$ MeV) by the formula (4a) exceeds up to about +20% for TlBr detector. It is considerably greater than for CdZnTe (Fig. 13) and HgI$_2$ (Fig. 15) detectors.

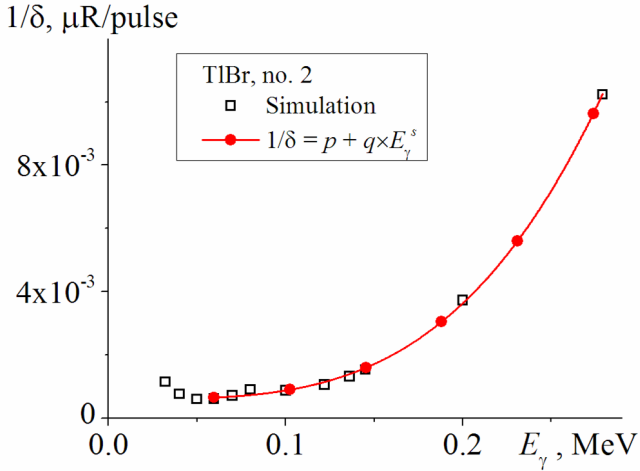

Fig. 16. The $1/\delta$ dependence at $E_\gamma$ below 0.3 MeV for TlBr detector.

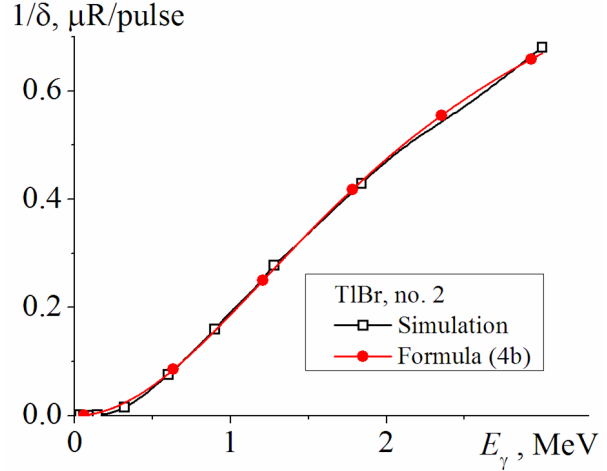

Fig. 17. The total $1/\delta$ dependence for TlBr detector.

The simulated TlBr detector has the least thickness among all investigated detectors (Table 1). Thus the simulation of the TlBr detector of 2 mm thickness with bias correspondingly increased up to 400 V was run to find out the reasons of so essential errors. From Fig. 19 it follows that for thicker TlBr detector the approximation error of $1/\delta$ is decreased in the energy range between 0.3 and 0.4 MeV similarly to error of $1/\delta$ for energy $E_\gamma$ equal 0.07 MeV. However the difference between the approximate formulas and the Monte-Carlo simulation is remained maximum for TlBr detectors in comparison with CdZnTe and HgI$_2$ detectors.

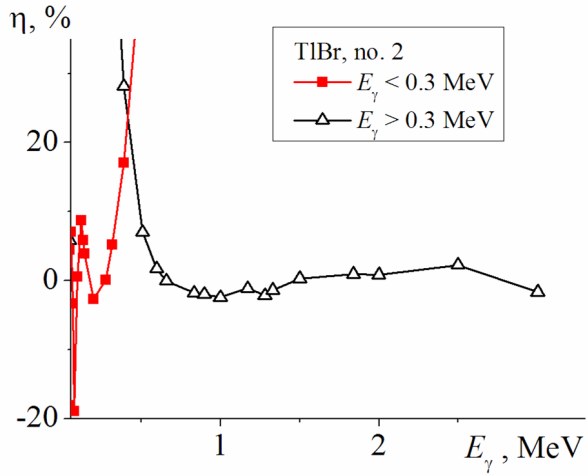

Fig. 18. The difference between the approximate $1/\delta$ calculation with formulas (4) and the Monte-Carlo simulation data for TlBr detector with the thickness of 1 mm.

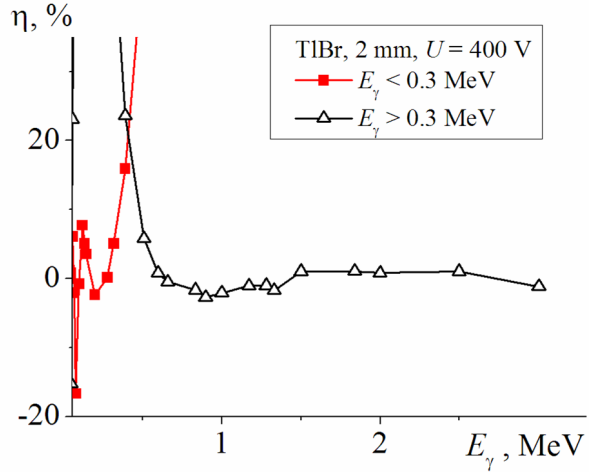

Fig. 19. The difference between the approximate $1/\delta$ calculation with formulas (4) and the Monte-Carlo simulation data for TlBr detector with the thickness of 2 mm.

Sensitivity $\delta(E_\gamma)$ of Ge gamma-radiation detector with characteristics corresponded to the GLP-36360/13P4 detector (ORTEC production [26]) was also calculated in order to compare that with results for detectors based on the wide band-gap semiconductor compounds. The difference between the simulated value of $1/\delta$ and the approximate formulas (4) is minimum among all materials investigated in this work (Fig. 20). The maximum error for Ge detector is about ±6% and it was found in the low energy range ($E_\gamma$ below 0.2 MeV) in contrast to the room-temperature semiconductor detectors.

Table 3 contains the calculated exponent values of the formulas (4) for all investigated semiconductors. From Table 3 it follows that for wide band-gap semiconductors $\frac{1}{\delta} \sim E_\gamma^{(3\ldots3,5)}$ in the low energy gamma-quantum region. The dependence of $1/\delta$ from $E_\gamma$ is square-law in this energy range for Ge detector. For wide band-gap semiconductors the sharper form of dependence of $1/\delta$ from $E_\gamma$ is probably connected with the jump increasing of the cross-section of full gamma-quantum absorption in the energy range between 0.06 and 0.3 MeV (Fig. 1). These jumps occur near the absorption *K*-edge of Hg, Tl, Cd, Te atoms. The binding energy of *K*-electrons in Ge is considerably less (about 11.1 keV) and thus the absorption coefficient changes more smoothly for this semiconductor.

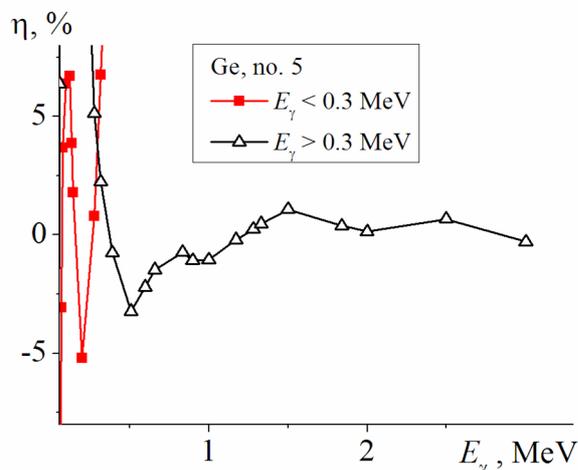

Fig. 20. The difference between the approximate $1/\delta$ calculation formulas (4) and the Monte-Carlo simulation data for Ge detector.

Table 3
The main fit parameters of the formulas (4)

| no. | SCD | $s$, (4a) | $1 - c$, (4b) |
|---|---|---|---|
| 1 | $HgI_2$ | 3.075±0.032 | 1+0.855 (±0.034) |
| 2 | TlBr | 3.45±0.11 | 1+0.96 (±0.05) |
| 3 | CdZnTe | 2.98±0.08 | 1+0.626 (±0.033) |
| 4 | CdZnTe | 3.46±0.06 | 1+0.618 (±0.032) |
| 5 | Ge | 2.07±0.12 | 1+0.367 (±0.019) |

It should be noted that for wide band-gap semiconductors the maximum difference between the fit formulas and the simulation results in the low energy range corresponds to the jumps of the total absorption cross-sections (between 70 and 80 keV).

The obtained formulas (4) are correct for SCD with thickness of more than 1 mm. As Fig. 2 shows the SCD sensitivity value in the gamma-quantum energy range $E_\gamma$ above 1 MeV is the same order with the noise level due to smaller thicknesses. Consequently the measurement of the sensitivity of the thin SCD in this energy range does not have the practical significance.

## CONCLUSIONS

The approximate formulas that allow to calculate the dependence of the semiconductor detector sensitivity $\delta(E_\gamma)$ from the energy of the registered gamma-quanta in the range between 0.06 and 3 MeV have been received. These formulas can be efficient when there is lack of experimental sensitivity measurements for more exact calculation. These approximate formulas can be also used instead of the detailed Monte-Carlo simulation of the $\delta(E_\gamma)$ dependence in the cases when the main control parameters of model (for example, the product of mobility and mean lifetime for electrons and holes and/or the detector internal electric field intensity) can not be determined with accuracy that is necessary for correct reconstruction of the SCD response functions.

It is shown that for the room-temperature detectors the maximum difference between the exact Monte-Carlo simulation and calculation with the approximate formulas is between 10 and 20% depending on the material. The region where maximum difference was found corresponds to the gamma-quantum energy range $E_\gamma$ between 0.3 and 0.4 MeV. The expected difference is from 2 to 4 times less in the range $E_\gamma$ below 0.3 MeV and in the range $E_\gamma$ above 0.3 MeV the expected difference is from 4 to 10 times less.

The obtained approximate formulas are correct for room-temperature semiconductor detectors with thickness of above 1 mm because the sufficiently great value of efficiency of the registration of the gamma-quanta with above 0.5 MeV energy can be received at this thickness range.

Further planned work related to the experimental verification of the obtained approximate formulas.